# Energy creation in a moving solenoid?

Nelson R. F. Braga[*] and Ranieri V. Nery[†]

*Instituto de Física, Universidade Federal do Rio de Janeiro,*

*Caixa Postal 68528, RJ 21941-972 – Brazil*

## Abstract

The electromagnetic energy $U_{em}$ stored in the magnetic field of an ideal solenoid at rest appears for a moving observer as $U'_{em}$, a sum of magnetic and electric energies satisfying: $U'_{em} > \gamma U_{em}$. We explain this seemingly paradoxical result calculating the stresses in the solenoid structure and showing that the total energy of the solenoid transforms by the expected relativistic factor $\gamma$.

[*]Electronic address: `braga@if.ufrj.br`

[†]Electronic address: `ranieri@ufrj.br`

## I. INTRODUCTION

We consider an infinitely long ideal solenoid with axis $z$ and radius $a$, as shown in fig. (1), at rest in an inertial frame $S_0$. There are $n$ turns of wire per unit length, carrying an electric current $i$. As is well known (e.g. [1]), the magnetic field $\vec{B}$ vanishes outside the solenoid and has a constant magnitude $B = \mu_0 n i$ inside the solenoid in the axial direction. We take the orientation of the current such that the field is in the positive $z$ direction.

Although this solenoid is infinite, there are neither external forces nor external energy fluxes acting on it. So it is an isolated system, for which we expect that the total energy should transform by a factor of $\gamma = 1/\sqrt{1 - v^2/c^2}$, when going to a frame where it is in motion with speed $v$.

Let us consider a frame $S'$, where the solenoid moves with constant velocity $v$ in the negative direction of the $x$ axis. Considering the relativistic transformations of the Electromagnetic fields, one finds that inside the solenoid there is a magnetic field $B' = \gamma B$, in the axial direction, and an electric field $E' = v\gamma B$, in the negative $y$ direction. Outside the solenoid the fields remain null. The internal volume of any finite portion of the solenoid is reduced by a factor of $1/\gamma$ because of Lorentz contraction in the $x$ direction.

In the rest frame $S_0$, the energy stored in the magnetic field in a volume $\Delta V = \pi a^2 \Delta z$ of the solenoid is

$$U_{em} = \frac{B^2}{2\mu_0} \Delta V. \tag{1}$$

In $S'$ the sum of the energies of the electric and magnetic fields is

$$\begin{aligned} U'_{em} &= \left(\frac{B'^2}{2\mu_0} + \frac{\varepsilon_0 E'^2}{2}\right) \frac{\Delta V}{\gamma} \\ &= \gamma \frac{B^2}{2\mu_0}\left(1 + \frac{v^2}{c^2}\right) \Delta V = \left(1 + \frac{v^2}{c^2}\right) \gamma\, U_{em}. \end{aligned} \tag{2}$$

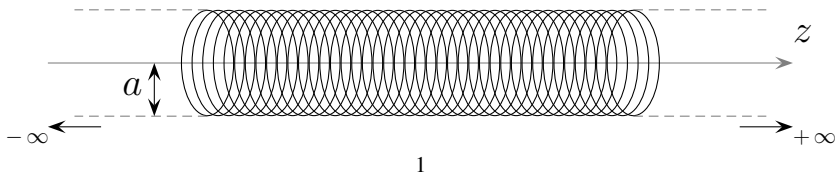

FIG. 1: Infinite solenoid

So, we find for the electromagnetic energy: $U'_{em} > \gamma U_{em}$. In order to understand why the electromagnetic energy does not transform with the factor $\gamma$, we must complete the analysis of the transformation of the energy, considering the fact that the solenoid is a continuous system. The energy density for continuous distributions of matter transforms as a component of a four-tensor that also includes the stresses[2], unlike the energy of point particles that transforms as a component of a four-vector. The energy density in the frame $S'$ includes a contribution due to the stresses in the rest frame $S_0$. This stress show up in the solenoid because the wire loops carry an electric current and are subject to a magnetic field.

We know the magnetic field inside and outside the wire loops (that means, respectively at $r < a$ and $r > a$). In order to determine the magnetic force, and the corresponding stress, on the wire, we must find the field at $r = a$ in the frame $S_0$. With this purpose, we can remodel the solenoid, taking the wires to have a finite width, meaning that the radius of the wire loops ranges from $r = a$ to $r = a + \Delta a$. The magnetic field in a point inside the wire at a radius $r$ corresponds to the field produced by the current situated at larger radii: $r' > r$. Increasing the radius by $dr$ we decrease this external current. Calling the decrease as $di$, the variation $dB$ of the magnetic field is

$$dB = -\mu_0 n di. \tag{3}$$

The force acting on an element of a wire with infinitesimal area $dr\, dl$ is

$$d^2 F = di B dl$$
$$= -\frac{B dB dl}{\mu_0 n}, \tag{4}$$

and points in the radial direction. Considering $dl = r d\theta$ and integrating by parts we have

$$dF = -\frac{1}{\mu_0 n} \left( \int_{B(a)}^{B(a+\Delta a)} B r dB \right) d\theta = -\frac{1}{\mu_0 n} \left( \int_{B^2(a)}^{B^2(a+\Delta a)} \frac{r}{2} dB^2 \right) d\theta$$
$$= -\frac{1}{\mu_0 n} \left( -\frac{1}{2} r B^2(r) \Big|_{r=a}^{r=a+\Delta a} - \int_a^{a+\Delta a} \frac{B^2}{2} dr \right) d\theta. \tag{5}$$

Taking now the limit of an ideal solenoid with a very thin wire: $\Delta a \to 0$, the term $\int_a^{a+\Delta a} \frac{B^2}{2} dr$ becomes negligible. Using the boundary conditions $B(a) = \mu_0 n i$ and $B(a + \Delta a) = 0$ the force over the infinitesimal element of the wire takes the form

$$dF = i \left( \frac{\mu_0 n i}{2} \right) a d\theta. \tag{6}$$

This is equivalent to the force produced by a magnetic field $\mu_0 n i/2$. So that we can take the magnitude of the field at $r = a$ as one half of the magnitude at $r < a$ (interior of the solenoid).

## II. CALCULATING THE STRESS

The magnetic force pushes all pieces of the solenoid in the positive radial direction. Considering that the solenoid is at equilibrium, this force must be balanced by the internal tensions in the wire. In order to simplify the calculations, we consider that our ideal wire has no width but it is surrounded by a cylindrical shell of width $\lambda$ made of a material that holds the stability of the system as shown in (fig. 2).

We calculate the stresses in this shell by considering a volume element inside it and using cylindrical coordinates. The symmetry of the solenoid implies that only the diagonal terms of the stress tensor are non vanishing. The equilibrium of forces leads to the following relation between the relevant components:

$$r\, t_{rr}(r)dzd\theta - [r + dr]\, t_{rr}(r + dr)dzd\theta + t_{\theta\theta}drdzd\theta = 0. \tag{7}$$

This implies:

$$\frac{d(t_{rr}r)}{dr} = t_{\theta\theta}. \tag{8}$$

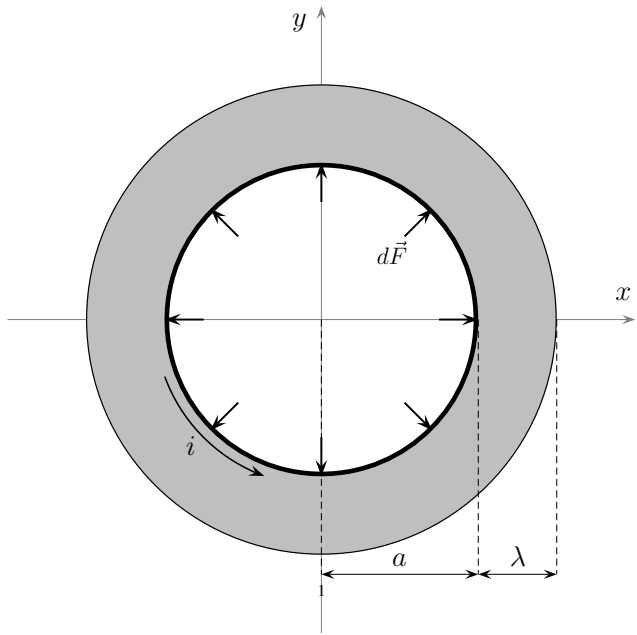

FIG. 2: Representation of the shell and the forces acting on the solenoid

The energy density for the matter content of the solenoid in $S'$ is given by $\gamma^2(\rho c^2 + t_{xx}(v^2/c^2))$, with $\rho$ being the mass density and $t_{xx}$ the stress along x and on a face orthogonal to this direction, both calculated in $S_0$. Since the volume of any part of the solenoid is reduced by an factor of $1/\gamma$, the first (mass) term transforms as expected, leaving the analysis only to the term involving the stress.

The relation between cartesian and cylindrical components is: $t_{xx} = t_{rr}\cos^2\theta + t_{\theta\theta}\sin^2\theta$. The extra contribution to the energy due to the stress is obtained integrating $\gamma^2 t_{xx}(v^2/c^2)$ along the cylindrical shell in the $S'$ frame. Considering a part of the shell with height $\Delta z$, that involves the volume $\Delta V$, we can integrate in $S'$ using the cylindric coordinates $(r,\theta)$ of $S_0$. Taking into account the contraction in the $x$ direction, the integral takes the form

$$\begin{aligned}\Delta U' &= \gamma^2 \frac{v^2}{c^2}\Delta z \int_a^{a+\lambda}\int_0^{2\pi} t_{xx}\frac{r}{\gamma}d\theta dr = \gamma\frac{v^2}{c^2}\pi\Delta z \int_a^{a+\lambda} rt_{rr}+rt_{\theta\theta}dr \\ &= \gamma\frac{v^2}{c^2}\pi\Delta z \int_a^{a+\lambda}\frac{d(r^2 t_{rr})}{dr}dr = -\gamma\frac{v^2}{c^2}\pi a^2 \Delta z t_{rr}(a) \end{aligned} \quad (9)$$

Where we used the fact that at the outside radius of the shell there are no external forces, so $t_{rr}(a+\lambda) = 0$. In order to calculate $t_{rr}$ in the inside radius of the shell we use eq. (6) and take a surface element $ad\theta\Delta z$, containing $n\Delta z$ spires to find

$$t_{rr}(a) = \frac{n\Delta z\, dF}{ad\theta\Delta z} = \frac{niB}{2} = \frac{B^2}{2\mu_0}, \quad (10)$$

so that

$$\Delta U' = -\gamma\frac{v^2}{c^2}\frac{B^2}{2\mu_0}\Delta V. \quad (11)$$

This is exactly the opposite of the exceeding term that appeared in the electromagnetic energy in $S'$. So

$$U'_{em} + \Delta U' = \gamma U_{em} \quad (12)$$

Showing that the whole solenoid indeed behaves as an isolated system. Similar problems were discussed for plane and spherical capacitors, respectively in [3] and [4].

**Acknowledgments:** The authors are partially supported by CNPq and FAPERJ.

---